\pdfoutput=1
\documentclass[printer]{tADR2e}
\usepackage{fixltx2e}
\usepackage{upgreek}
\usepackage{amsmath}
\usepackage{physics}
\usepackage{hyperref}

\def\SPSB#1#2{\rlap{\textsuperscript{{#1}}}\SB{#2}}

\def\SB#1{\textsubscript{{#1}}}

\begin{document}

\jvol{31} \jnum{17} \jyear{2017} \jmonth{September}

\articletype{FULL PAPER}

\title{Mechanical principles of dynamic terrestrial self-righting using wings}

\author{Chen Li$^{a}$$^{\ast}$\thanks{$^\ast$Corresponding author. Email: chen.li@jhu.edu. Website: https://li.me.jhu.edu
\newline{Portions of this work were previously presented at the 2016 IEEE/RSJ International Conference on Intelligent Robots and Systems (IROS 2016), Daejeon, Korea~\cite{Li2016cockroach}.}}, Chad C. Kessens$^{b}$, Ronald S. Fearing$^{c}$, and Robert J. Full$^{c}$\\\vspace{6pt}  $^{a}${\em{Johns Hopkins University, Baltimore, MD 21218, USA}};\\
$^{b}${\em{Army Research Laboratory, Aberdeen Proving Ground, MD 21005, USA}};\\
$^{c}${\em{University of California, Berkeley, CA 94720, USA}}\\
\vspace{6pt}\received{Received 17 January, 2017; accepted DATE MONTH, 2017}}

\maketitle

\begin{abstract}

Terrestrial animals and robots are susceptible to flipping-over during rapid locomotion in complex terrains. However, small robots are less capable of self-righting from an upside-down orientation compared to small animals like insects. Inspired by the winged discoid cockroach, we designed a new robot that opens its wings to self-right by pushing against the ground. We used this robot to systematically test how self-righting performance depends on wing opening magnitude, speed, and asymmetry, and modeled how kinematic and energetic requirements depend on wing shape and body/wing mass distribution. We discovered that the robot self-rights dynamically using kinetic energy to overcome potential energy barriers, that larger and faster symmetric wing opening increases self-righting performance, and that opening wings asymmetrically increases righting probability when wing opening is small. Our results suggested that the discoid cockroach's winged self-righting is a dynamic maneuver. While the thin, lightweight wings of the discoid cockroach and our robot are energetically sub-optimal for self-righting compared to tall, heavy ones, their ability to open wings saves them substantial energy compared to if they had static shells. Analogous to biological exaptations, our study provided a proof-of-concept for terrestrial robots to use existing morphology in novel ways to overcome new locomotor challenges.
%\medskip

%\medskip

\end{abstract}

\section{INTRODUCTION}

Mobile robots have begun to venture out of the laboratory and into the real-world~\cite{brooks2004robots} and are anticipated to impact a broad range of scenarios important to society~\cite{Siciliano2016handbook}, such as search and rescue~\cite{Saranli2001rhex,murphy2010trial,Huang2011operation,Wood2013flight,Low2015Perspectives}, precision agriculture~\cite{Astrand2002agricultural,Billingsley2008robotics,Wood2013flight,Emmi2013Fleets}, environmental monitoring~\cite{Lauder2007fish,Moored2011Batoid,Wood2013flight,Low2015Perspectives}, structure examination~\cite{Kim2008smooth,Rollinson2014pipe}, public safety~\cite{Trevelyan1997robots,Zeng2007research}, and extra-terrestrial exploration~\cite{matson2010unfree,Bostock2014month}. To complete these tasks, mobile robots must be able to locomote through a diversity of complex terrains ranging from desert sand~\cite{li2009sensitive,Roberts2014deserta}, loose soil~\cite{Yang2011experimental}, cluttered vegetation and foilage~\cite{McKenna2008Toroidal,Li2015terradynamically}, to building rubble~\cite{murphy2010trial} and Martian soil~\cite{matson2010unfree,Bostock2014month}, which can often be uneven~\cite{Saranli2001rhex,Kim2007Walking,Raibert2008BigDog}, sloped~\cite{Bares1999dante,Chew1999Blind,Kim2007Walking,Peters2008Mobile}, dispersed~\cite{spagna2007distributed}, cluttered~\cite{McKenna2008Toroidal,Pettersen2011snake,Li2015terradynamically,Travers2016Shape}, or even flowable~\cite{li2013terradynamics,li2009sensitive,Roberts2014deserta,Yang2011experimental,Gong2015Simplifying}. Locomotion on such challenging terrains can not only induce static and dynamic instability and translational and rotational perturbations~\cite{Saranli2001rhex,Li2015terradynamically,Bermudez2012Performance}, but also cause the robot to suffer from loss of foothold~\cite{spagna2007distributed,li2009sensitive} and an inability to generate appropriate ground reaction forces~\cite{li2009sensitive,matson2010unfree,Roberts2014deserta,Yang2011experimental,Gong2015Simplifying}. All these could lead to the robot flipping-over and losing mobility~\cite{Guizzo2015hard}. Terrestrial self-righting from an upside-down orientation is therefore a critical locomotor capability for mobile robots to ensure continuous operation.

A diversity of terrestrial self-righting techniques has been developed to help mobile robots self-right. These include: having a body shape that is unstable when upside down together with a low or movable center of mass position ~\cite{Brown1998bow,Fiorini2003development,Aoyama2005Micro,wallace2007Biologically,Kovac2009steerable,Beyer2009performance}; implementing additional long appendages such as arms, levers, legs, or tails~\cite{Fiorini2003development,Kessens2013Framework,Kossett2011robust,Klaptocz2012Active,Zhang2012self,Briod2012AirBurr,Krummel2014horseshoe,peng2015motion}; using reconfigurable wheels~\cite{Tunstel1999evolution}, tracks~\cite{Schempf1999pandora,Zong2006Realization}, or body modules~\cite{Yim2007towards,Zong2006Realization} that can be re-configured via self-reassembly to change overall shape; or working around the problem of flipping-over by adopting a dorsoventrally symmetrical body design~\cite{Saranli2004model,Ben-Tzvi2008Design,Kossett2011robust} or one with no ``upright'' orientation if a nominal upright orientation is not required~\cite{Tsukagoshi2005Design}.

Rapidly-running small legged robots such as RHex~\cite{Saranli2001rhex}, iSprawl~\cite{kim2006isprawl}, and VelociRoACH~\cite{Haldane2013Animal} are particularly easy to flip over in complex terrains, because their small body inertia combined with terrain irregularities that are comparable to their size~\cite{Kaspari1999size} can lead to large perturbations. However, to ensure running capacity and dynamic stability, these robots usually have multiple legs that are short relative to body size and compliant enough to use elastic energy-saving mechanisms during dynamic running~\cite{blickhan1989spring}. As a result, their legs are often not directly useful for terrestrial self-righting, or it requires careful motion planning to do so (for example, RHex can use a series of carefully planned ground impact to store elastic energy in its legs and then release it to perform aerobatics to self-right~\cite{Saranli2004model,Saranli2004multipoint,Johnson2013legged}). While many of the self-righting techniques mentioned above have been successfully demonstrated or implemented in other platforms, few of them have been used on rapidly-running small legged robots partly due to their limited payload.

\begin{figure}[thpb]
      \centering
      \includegraphics[scale=0.79]{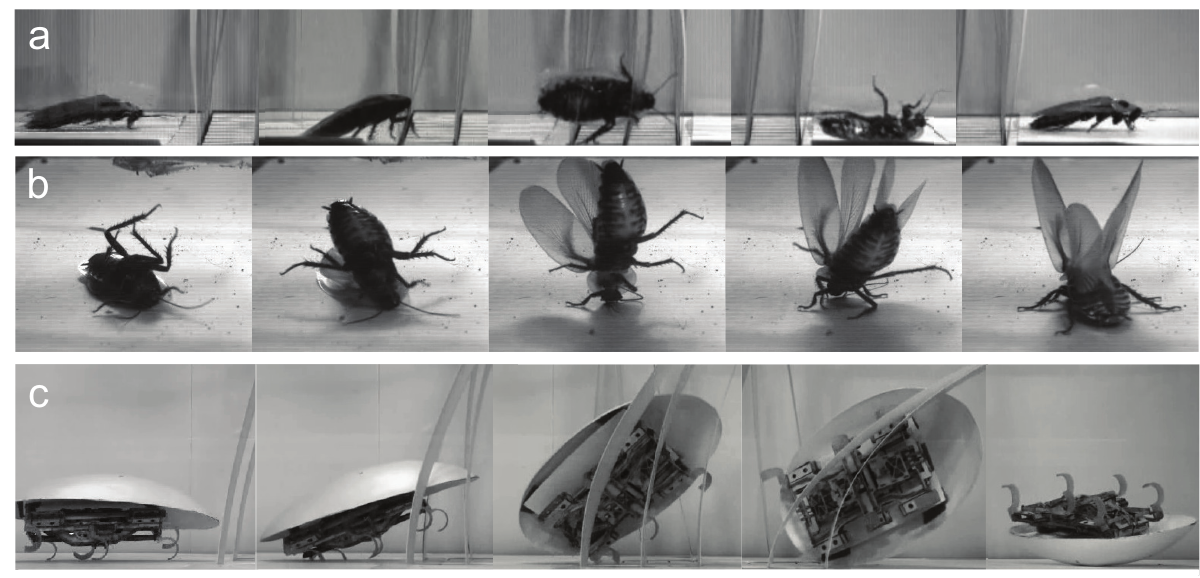}
      \caption{
      Small insects like cockroaches use \textbf{exaptations} of its wings to facilitate obstacle traversal and self-right from an upside-down orientation, providing inspiration for small legged robots to achieve such multi-functional locomotion~\cite{Low2015Perspectives} in complex terrain. (a) A discoid cockroach rapidly traverses cluttered obstacles such as grass-like beams~\cite{Li2015terradynamically}, during which its wings are folded against the body as a rounded ellipsoidal ``shell'' to facilitate body rolling; after flipping-over, it quickly rights itself. (b) The cockroach self-rights by opening and pushing its wings against the ground~\cite{li2015fast}. (c) A small legged robot uses a cockroach-inspired rounded shell to traverse cluttered obstacles; however, when it over-rolls and flips over, it becomes stable and cannot self-right~\cite{Li2015terradynamically}.
      }
      \label{motivation}
\end{figure}

Small animals like insects face similar challenges of flipping-over as small robots do~\cite{Kaspari1999size}. To self-right after flipping-over, many small animals use \textbf{exaptations}~\cite{Jay1982exaptation} of appendages that primarily serve other purposes~\cite{Brackenbury1990novel,full1995maximum,Frantsevich2004righting,Young2006effects,Domokos2008Geometry}. For example, the discoid cockroach, \textit{Blaberus discoidalis}, is a flightless insect that lives in cluttered forest floor and moves through cluttered foilage and litter on a daily basis. Normally, its wings are folded against the body, forming a protective ``shell''. This ``shell'' also helps the animal traverse cluttered obstacles such as grass-like beams, because its rounded shape is ``terradynamically streamlined'' and reduces terrain resistance by facilitating body rolling~\cite{Li2015terradynamically} (Figure~\ref{motivation}a). However, when flipped over (Figure~\ref{motivation}a), the discoid cockroach can also open its wings and use them to rapidly push against the ground to right itself (Figure~\ref{motivation}b)~\cite{li2015fast}.

We were greatly inspired by the discoid cockroach's remarkable ability to use the same body structures for multi-functional locomotion~\cite{Low2015Perspectives}. In a recent study, we first enabled a small legged robot to traverse cluttered obstacles by adding a ``terradynamically streamlined'' rounded ellipsoidal shell~\cite{Li2015terradynamically} (Figure~\ref{motivation}c). However, the robot still lacked the animal's multi-functional abilities: when it over-rolled during obstacle traversal, the robot became permanently flipped over because the rounded shell is stable when upside down (Figure~\ref{motivation}c)~\cite{Li2015terradynamically}.

In this study, we take the next step towards multi-functional locomotion~\cite{Low2015Perspectives} for small legged robots using \textbf{exaptations} by further developing the rounded shell into actuated wings to enable self-righting capability. We performed systematic experiments to study how winged self-righting performance depends on the speed and magnitude of wing opening, and used a simple dynamic model to understand the falling phase of self-righting. We then applied a planar geometric modeling framework~\cite{Kessens2013Framework,Kessens2014metric} to study how the kinematic and energetic requirements for dynamic self-righting depend on wing shape and body/wing mass distribution, and determined whether quasi-static righting is possible for the robot and the discoid cockroach. Finally, inspired by observations from cockroach winged self-righting experiments, we performed robot experiments by using different left and right wing opening to understand whether asymmetric wing opening provides any advantage for self-righting~\cite{li2015fast}.

\section{WINGED SELF-RIGHTING MECHANISM DESIGN}

\begin{figure}[thpb]
      \centering
      \includegraphics[scale=0.77]{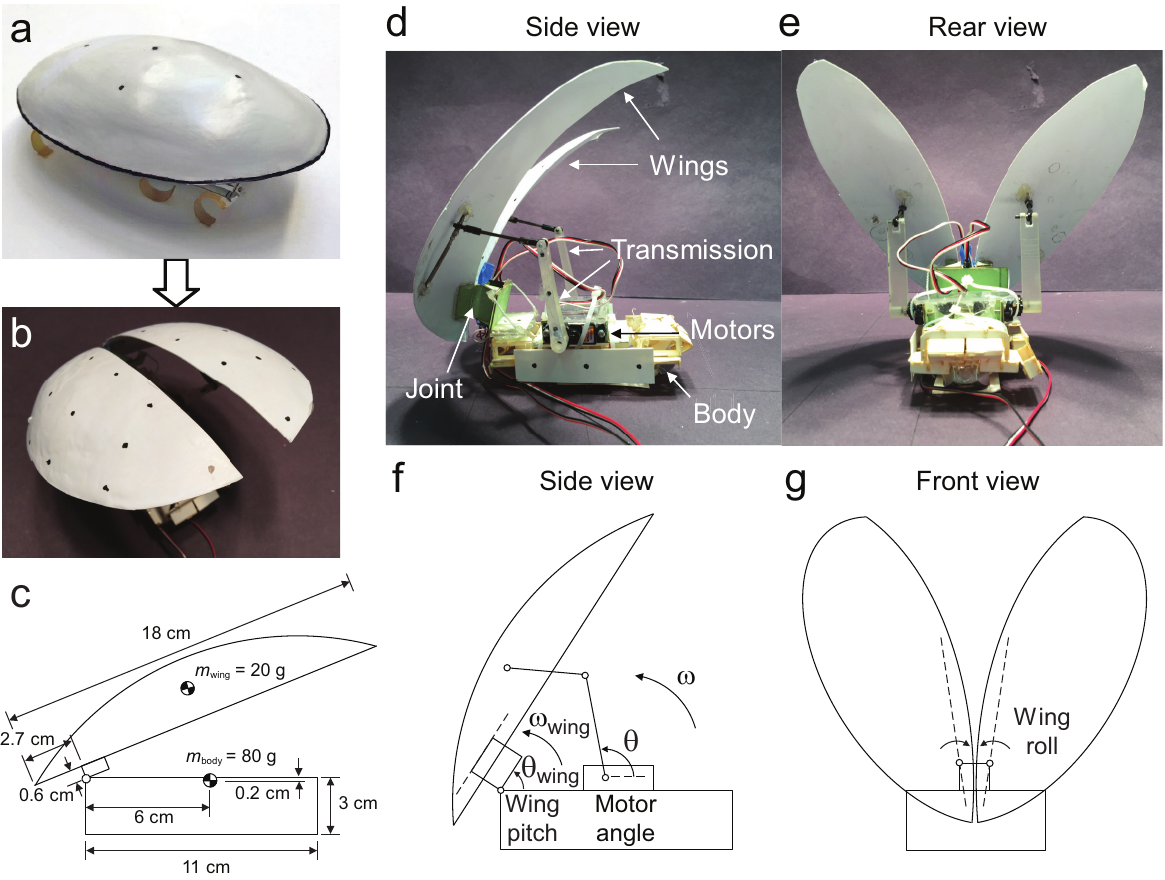}
      \caption{
      Design of winged self-righting mechanism. (a, b) The ``terradynamically streamlined'' rounded shell previously developed for obstacle traversal was transformed into two actuated wings for self-righting. (c) Sagittal plane schematic of the robot with geometric dimensions and mass distribution. (d, e) Side and rear views of the actuated wings mounted on a VelociRoACH robot body. (f) Side view design diagram of the pitching degree of freedom of the wings enabled by a four bar linkage transmission. (g) Front view design diagram of the rolling degree of freedom of the wings.
      }
      \label{design}
\end{figure}

Our winged self-righting mechanism was inspired by the discoid cockroach's multi-functional wings (Figure~\ref{design}). With future integration of self-righting and obstacle traversal~\cite{Li2015terradynamically} capabilities in mind, we formed the two actuated wings by sagittally slicing the same rounded ellipsoidal shell previously developed for obstacle traversal (Figure~\ref{design}a, b), adding actuators and transmissions for each half. We measured geometric dimensions and mass properties of the robot from the physical robot and its CAD model for use as model parameters (Figure~\ref{design}c). Technical details of the winged robot development were reported in detail in~\cite{Li2016cockroach}; below, we briefly review it to help understand experimental and modeling results.

The wings were attached to the anterior end of the robot body. When folded against the body, the two wings formed a rounded shell, similar to the discoid cockroach (Figure~\ref{design}b). When actuated, the wings could open in a similar fashion as the discoid cockroach (Figure~\ref{design}d, e), with both pitching (Figure~\ref{design}f) and rolling motion (Figure~\ref{design}g) relative to the body. The two wings were actuated by two small, lightweight, high-torque servo motors (Hyperion DS11-AMB), and could be either opened in exactly the same way (symmetric wing opening) or differently (asymmetric wing opening).

Because our study focused on discovering the principles of terrestrial winged self-righting, we tested the winged self-righting mechanism on a legless VelociRoACH robot body. In addition, we used an external power supply to provide constant power so that motor performance did not decrease with draining battery. Further, to minimize cable drag force during self-righting and interference with wing actuation, we used fine cables for power and control signals and carefully routed them through a small gap between the two wings at the anterior end of the robot.

\section{SYMMETRIC RIGHTING EXPERIMENTS}

\subsection{Body rotation during successful \& failed righting}

\begin{figure}[bthp]
      \centering
      \includegraphics[scale=0.86]{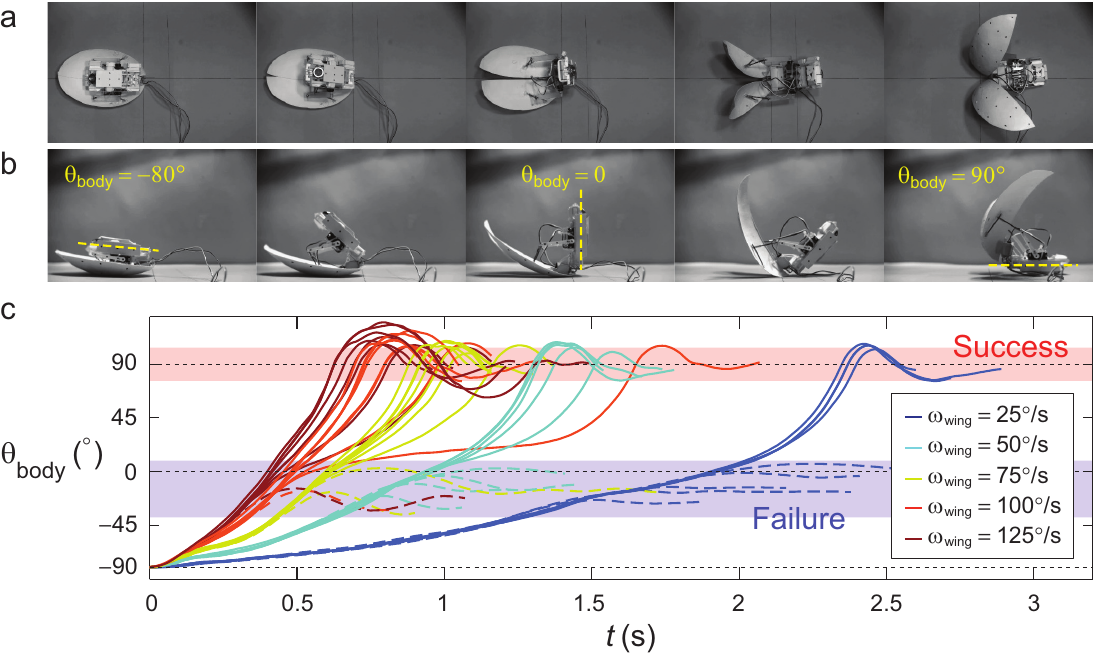}
      \caption{
      The robot's body rotation during self-righting using symmetric wing opening. (a, b) Top and side views of the robot during a successful self-righting maneuver at a few representative instances, with definition of body pitch angle, $\theta\textsubscript{body}$. (c) $\theta1\textsubscript{body}$ as a function of time, $t$, for a wide range of wing opening speeds $\omega\textsubscript{wing}$ and angle $\theta\textsubscript{wing}$ tested. Each curve of the same color uses the same $\omega\textsubscript{wing}$ (see legend), but different $\theta\textsubscript{wing}$. Solid and dashed curves represent successful and failed maneuvers, respectively. One representative trial was shown for each combination of $\theta\textsubscript{wing}$ and $\omega\textsubscript{wing}$.
      }
      \label{righting}
\end{figure}

To begin to discover the principles of terrestrial self-righting using wings, we performed robot experiments on a level, flat, rigid ground. We set up two synchronized high-speed cameras to record the robot's self-righting movement from both top and side views  (Figure~\ref{righting}a, b) at 100 frame$\cdot$sec$^{-1}$, and tracked markers positioned on the robot to measure its kinematics. Before each trial, we placed the robot upside down with the wings fully folded against the body and carefully positioned it at the same location and orientation in the camera views. We controlled the two wings to open in synchrony (symmetric wing opening) with the same magnitude and speed, and used the tracking data to measure the robot's body pitch angle, $\theta\textsubscript{body}$, as a function of time (Figure~\ref{righting}c).

Similar to the discoid cockroach, as the robot opened and pushed its wings against the ground, its body pitched up and rotated about the ground contact. As body pitching continued, the robot body eventually vaulted over the front edge of the wings and fell to the ground in an upright orientation, resulting in successfully righting. In this process, body pitch angle $\theta\textsubscript{body}$ changed from an initial $-$80$^\circ$ to a final 90$^\circ$ (Figure~\ref{righting}c). The body often oscillated as it impacted the ground before settling into a stationary upright orientation. Because we did not control the wings to fold back against the body after opening, failed righting maneuvers resulted in the body settling into a nearly vertical stationary orientation with $\theta\textsubscript{body} \approx$ 0.

\subsection{Effect of wing opening magnitude \& speed on righting performance}

\begin{figure}[t]
      \centering
      \begin{tabular}
        {@{}ccc@{}}
            \includegraphics[scale=0.4]{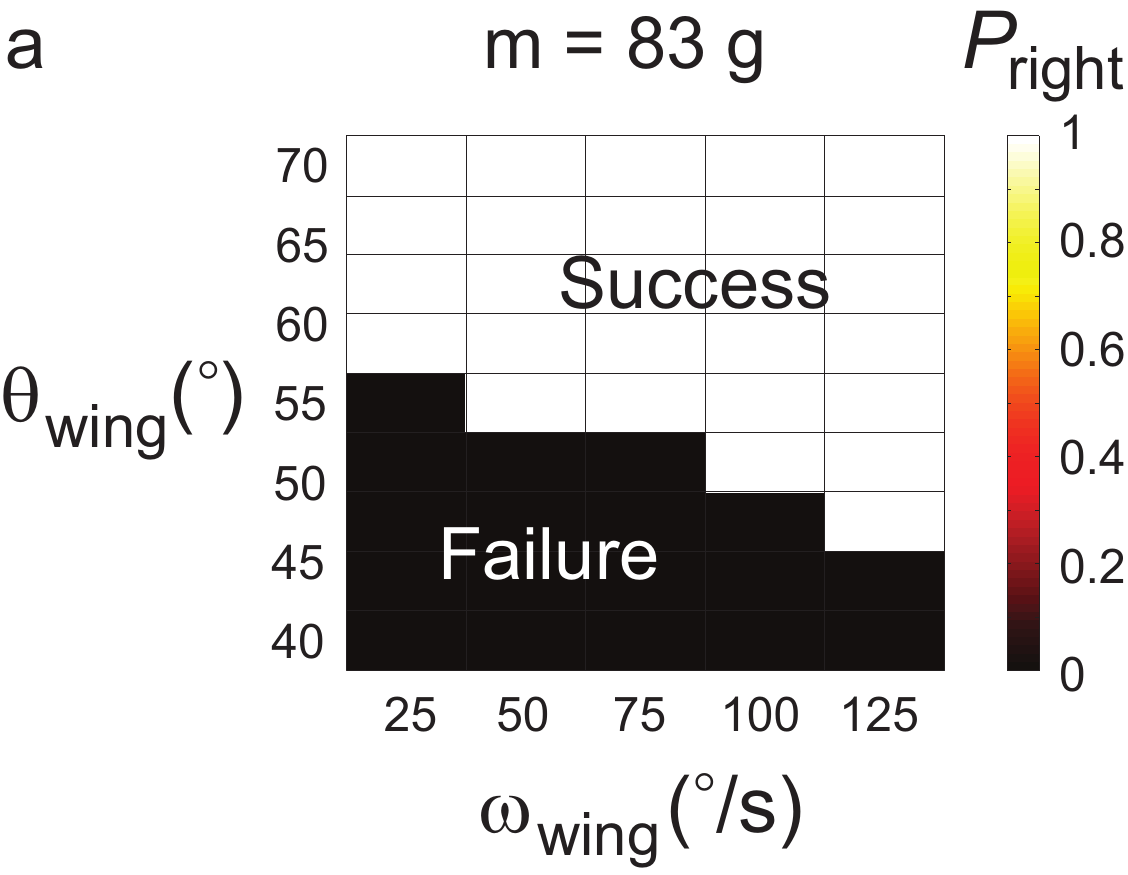} &
            \includegraphics[scale=0.4]{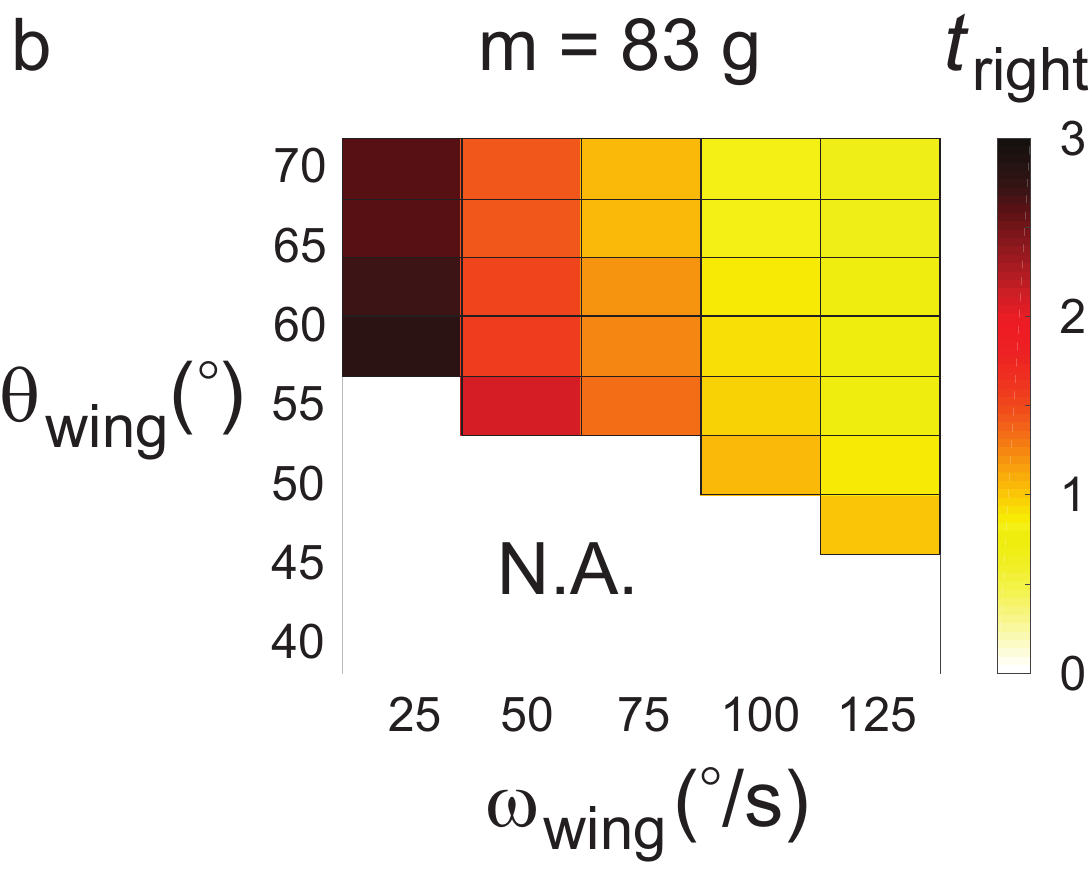} \\
            \includegraphics[scale=0.95]{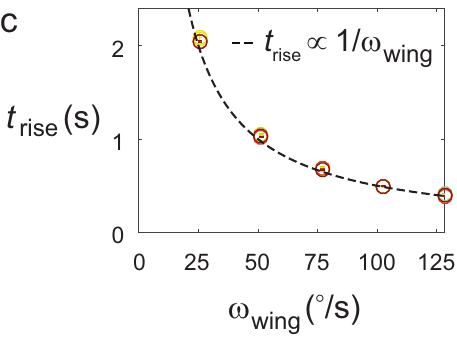} &
            \includegraphics[scale=0.95]{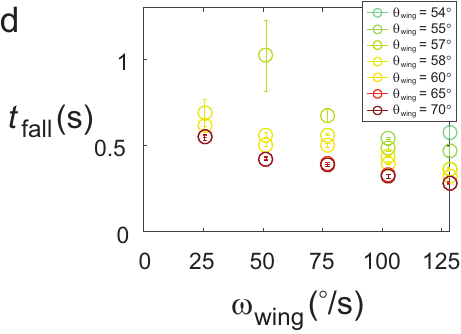} \\
      \end{tabular}
      \caption{
      Dependence of self-righting performance of the robot on the magnitude and speed of wing opening, for robot mass $m = 83$ g. (a) Righting probability, $P\textsubscript{right}$ (success = 1, failure = 0), and (b) righting time, $t\textsubscript{right}$, as a function of wing opening magnitude $\theta\textsubscript{wing}$ and speed $\omega\textsubscript{wing}$. Brighter colors represent higher performance and darker colors represent lower performance in (a) and (b). Color bar is used for $P\textsubscript{right}$ because average probability in principle can be any value between 0 and 1. (c) rising time, $t\textsubscript{rise}$, and (d) falling time, $t\textsubscript{fall}$, as a function of $\omega\textsubscript{wing}$ for a wide range of $\theta\textsubscript{wing}$. Redder colors represent higher $\theta\textsubscript{wing}$ and bluer colors represent lower $\theta\textsubscript{wing}$ in (c) and (d). Dashed curve in (c) is $t\textsubscript{rise} \propto 1/\omega$.
      }
      \label{performance}
\end{figure}

To understand how wing motion contributes to self-righting performance, we systematically tested the robot's righting probability, $P\textsubscript{right}$ (success = 1, failure = 0), and righting time, $t\textsubscript{right}$, depended on wing opening magnitude and speed (Figure~\ref{performance}) . Although wing opening involved both pitching and rolling motions of the wings relative to the body, for simplicity we used wing pitch angle $\theta\textsubscript{wing}$ and wing pitch angular velocity $\omega\textsubscript{wing}$ to represent wing opening magnitude and wing opening speed\footnote{Note that in the previous paper where we reported portions of the results, motor angle $\theta$ and motor speed $\omega$ were used as a proxy. Calibration showed that $\theta\textsubscript{wing} \approx \theta/2$ and $\omega\textsubscript{wing} \approx \omega/2$. Therefore, the data reported in the pervious paper and the present paper are consistent.}. Our calibration showed that it only took $\approx$ 10\% of the actuation time for the wings to accelerate and decelerate to the desired speeds, so we simply used the average angular velocity assuming instantaneous acceleration and deceleration. We verified that the motors were strong enough to overcome the weight of the robot in an upside-down orientation and accurately achieved the commanded actuation positions and speeds for the wing opening magnitudes and speeds tested, 40$^\circ \leq \theta \leq$ 70$^\circ$, 25$^\circ$/s $\leq \omega \leq$ 125$^\circ$/s. Three trials were performed for each combination of $\theta\textsubscript{wing}$ and $\omega\textsubscript{wing}$.

We observed that the robot's self-righting probability increased with both wing opening magnitude and wing opening speed (Figure~\ref{performance}a). For large wing opening magnitudes ($\theta>$ 55$^\circ$), the robot always righted ($P\textsubscript{right}$ = 1), even when wing opening was the slowest tested ($\omega\textsubscript{wing}$ = 25$^\circ$/s). For small wing opening magnitudes ($\theta <$ 44$^\circ$), the robot always failed to right ($P\textsubscript{right}$ = 0), even when wing opening was the fastest tested ($\omega\textsubscript{wing}$ = 125$^\circ$/s). For intermediate wing opening (44$^\circ \leq \theta \leq$ 55$^\circ$), the robot righted when wing opening was faster (larger $\omega\textsubscript{wing}$), but failed when wing opening was slower (smaller $\omega\textsubscript{wing}$). In addition, the robot self-righted more quickly as wing opening became larger or faster (Figure~\ref{performance}b): $t\textsubscript{right}$ decreased with $\omega\textsubscript{wing}$ for any given $\theta\textsubscript{wing}$ and decreased with $\theta\textsubscript{wing}$ for any given $\omega\textsubscript{wing}$.

To better understand the righting process, we further divided successful righting trials into a rising phase and a falling phase, and measured the time for each phase. Rising time $t\textsubscript{rise}$  was the time it took for the robot body to rotate from the initial upside-down orientation to a vertical orientation. Falling time $t\textsubscript{fall}$ was the subsequent time for the robot body to rotate from a vertical orientation to the final upright orientation. We observed that $t\textsubscript{rise}$ was insensitive to $\theta\textsubscript{wing}$ and nearly inversely proportional to $\omega\textsubscript{wing}$ (Figure~\ref{performance}c), suggesting that during the rising phase the robot moved in a kinematically similar fashion (only proportionally more rapidly) as wing opening speed increased. By contrast, $t\textsubscript{fall}$ not only decreased with $\omega\textsubscript{wing}$ for any given $\theta\textsubscript{wing}$, but also decreased with $\theta\textsubscript{wing}$ for any given $\omega\textsubscript{wing}$ (Figure~\ref{performance}d).

\subsection{Effect of body/wing mass distribution}

Further, to understand how body/wing mass distribution affected self-righting performance, we performed the same experiments and analysis using the robot with an additional mass on the robot. The total mass $m$ was increased from 83 g to 100 g, and the percentage of mass in the wings decreased from 25\% to 20\%. Because we observed excellent repeatability for nearly all $\theta\textsubscript{wing}$ and $\omega\textsubscript{wing}$ tested (standard deviation of $t\textsubscript{right} <$ 0.1 s) during experiments with a smaller total mass ($m$ = 83 g), in the experiments with a larger total mass ($m$ = 100 g), we performed one trial for each combination of $\theta\textsubscript{wing}$ and $\omega\textsubscript{wing}$.

\begin{figure}[t]
      \centering
      \begin{tabular}
        {@{}ccc@{}}
            \includegraphics[scale=0.4]{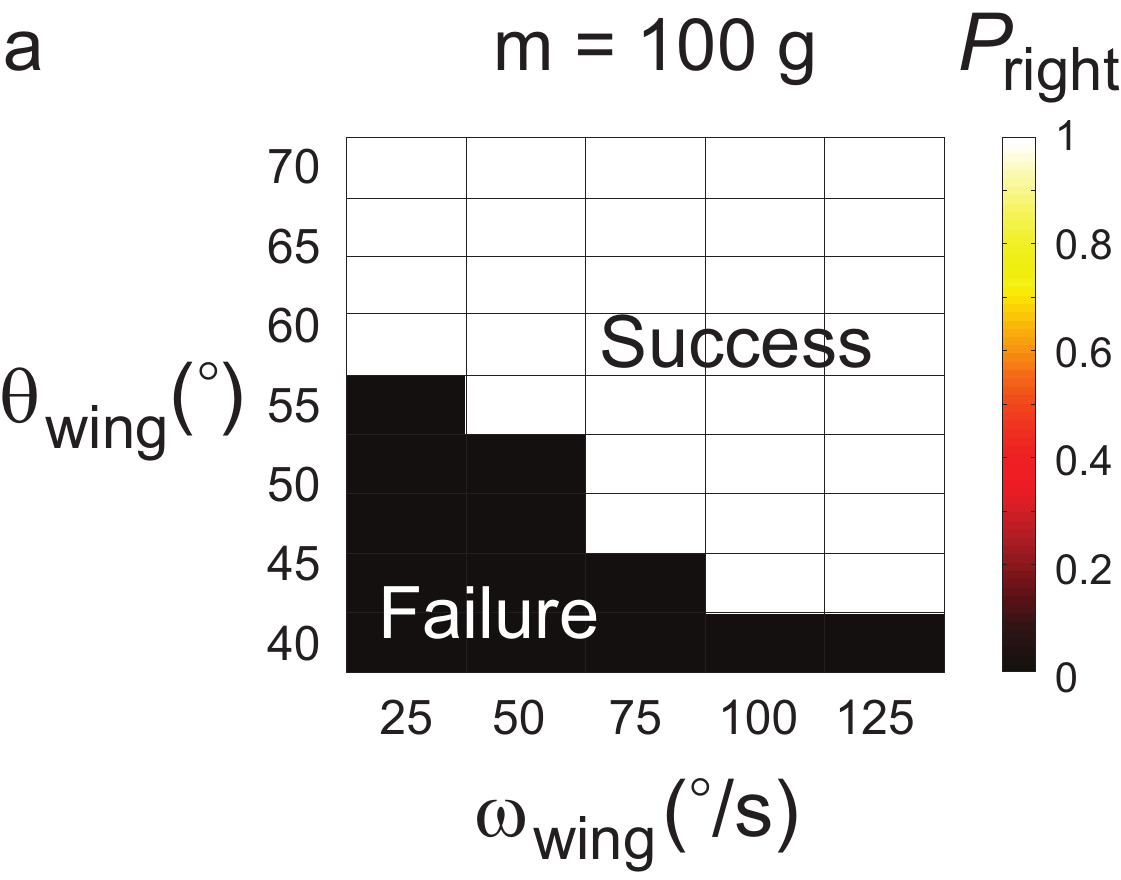} &
            \includegraphics[scale=0.4]{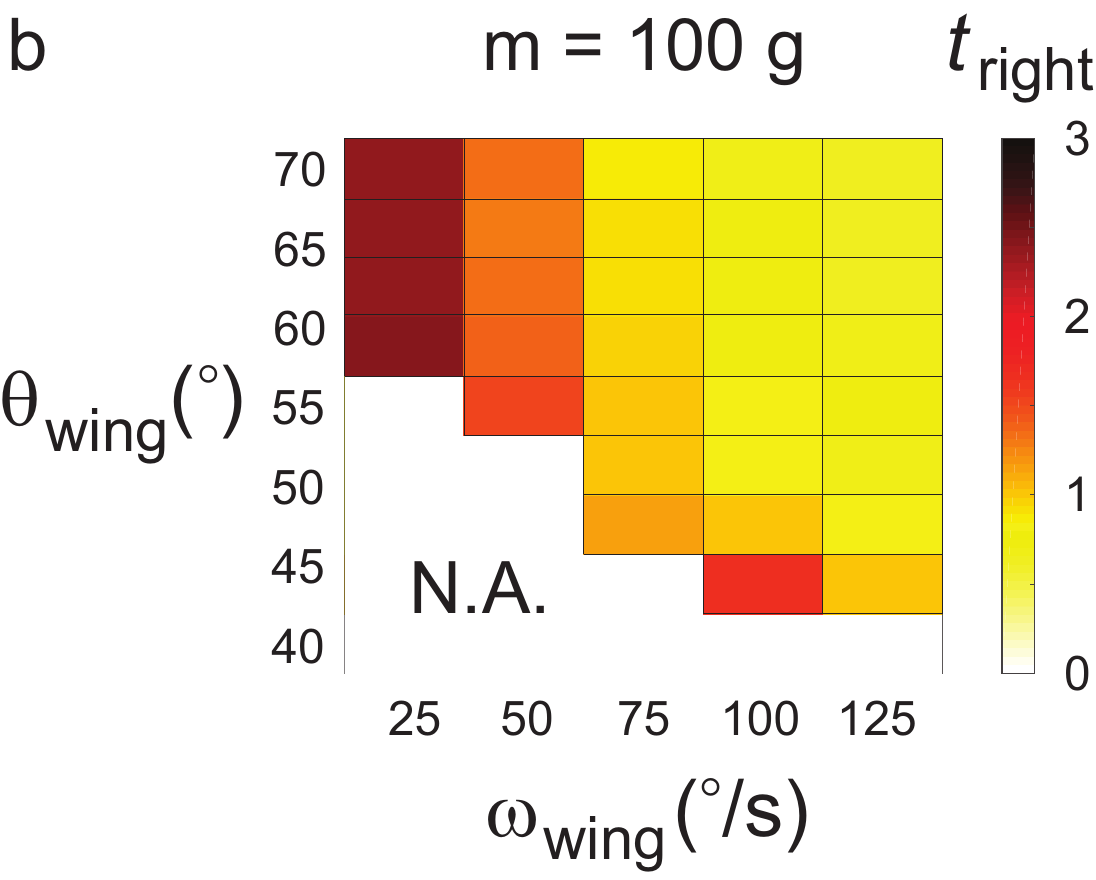} \\
            \includegraphics[scale=0.95]{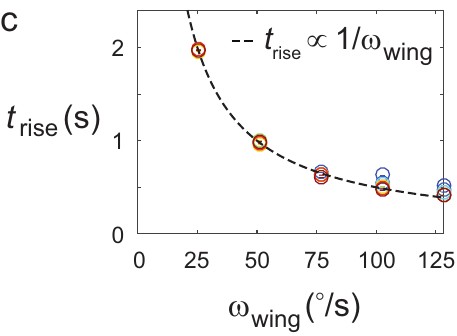} &
            \includegraphics[scale=0.95]{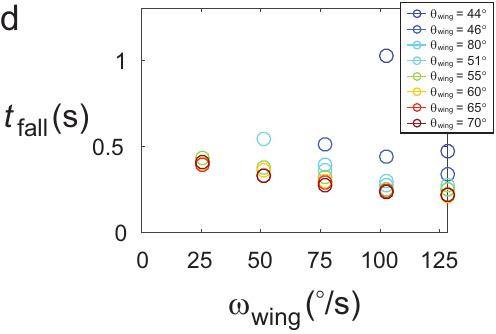} \\
      \end{tabular}
      \caption{
      Dependence of self-righting performance of the robot on the magnitude and speed of wing opening, for robot mass $m = 100$ g. (a) Righting probability, $P\textsubscript{right}$ (success = 1, failure = 0), and (b) righting time, $t\textsubscript{right}$, as a function of wing opening magnitude $\theta\textsubscript{wing}$ and speed $\omega\textsubscript{wing}$. Brighter colors represent higher performance and darker colors represent lower performance in (a) and (b). Color bar is used for $P\textsubscript{right}$ because average probability in principle can be any value between 0 and 1. (c) rising time, $t\textsubscript{rise}$, and (d) falling time, $t\textsubscript{fall}$, as a function of $\omega\textsubscript{wing}$ for a wide range of $\theta\textsubscript{wing}$. Redder colors represent higher $\theta\textsubscript{wing}$ and bluer colors represent lower $\theta\textsubscript{wing}$ in (c) and (d). Dashed curve in (c) is $t\textsubscript{rise} \propto 1/\omega$.
      }
      \label{performanceweight}
\end{figure}

Qualitatively, the robot with less mass distributed in the wings behaved the same as that with more mass on the wings (Figure~\ref{performanceweight}). However, the robot was able to successfully self-right at lower $\omega\textsubscript{wing}$ for any given $\theta\textsubscript{wing}$, and at lower $\theta\textsubscript{wing}$ for any given $\omega\textsubscript{wing}$ (Figure~\ref{performanceweight}a), and righting time was also slightly reduced for each combination of $\theta\textsubscript{wing}$ and $\omega\textsubscript{wing}$ (Figure~\ref{performanceweight}b). Rising and falling time showed similar dependence on $\theta\textsubscript{wing}$ and $\omega\textsubscript{wing}$ as those with more mass distributed in the wings (Figure~\ref{performanceweight}c, d).

\section{SYMMETRIC RIGHTING DYNAMIC MODELING}

\subsection{Equation of motion}

When self-righting using wings, the discoid cockroach held its wings nearly stationary relative to its body during the falling phase, and its legs did not contact the ground until the very end of the falling phase when body orientation was nearly upright (Figure~\ref{motivation}a)~\cite{li2015fast}. In our robot experiments, wings were similarly held stationary relative to the body after wing opening (Figure~\ref{righting}b), and legs were absent. In addition, we observed that the robot body rotation was confined within the sagittal plane when wing opening was symmetric, and that there was no slip and little rolling where the robot wings contacted the ground. Based on these observations, we treated the robot as a rigid body rotating in the sagittal plane about a fixed pivot on the ground under the torque of gravity, and developed a planar dynamic model to describe the passive falling dynamics of winged self-righting (Figure~\ref{dynamics}a).

Using the Lagrangian method, we derived the equation of motion as:
\begin{equation}
\dv{\omega\textsubscript{body}}{t} - \frac{mgL}{I\textsubscript{pitch} + mL^2} \textrm{sin}\theta\textsubscript{body} = 0
\end{equation}
where $\theta\textsubscript{body}$ is the body pitch angle, $\omega\textsubscript{body} = d\theta\textsubscript{body}/{dt}$ is the body pitch angular velocity, $t$ is time ($t$ = 0 at the beginning of passive falling), $m$ = 0.083 or 0.1 kg is the total mass of the robot, $L \approx$ 0.08 m is the distance between the center of mass and the ground contact point (front edge of the wings), $I\textsubscript{pitch} \approx$ 1.1 $\times$ 10$^{-4}$ kg m$^2$ is the moment of inertia of the robot about the center of mass along the pitch axis, and $g$ = 9.81 m/s$^2$ is the gravitational acceleration. The $mL^2$ term added to $I\textsubscript{pitch}$ is using the parallel axis theorem. We measured $L$ and $I\textsubscript{pitch}$ from a CAD model of the robot and verified that $L$ and $I\textsubscript{pitch} + mL^2$ changed little (up to $\pm$15\%) as the wings open. Thus, we assumed a constant value for $L$ and $I\textsubscript{pitch} + mL^2$. Using equation (1), we applied the Euler method to numerically integrate forward in time and calculated $\omega\textsubscript{body}(t)$ and $\theta\textsubscript{body}(t)$ given initial conditions $\omega\textsubscript{body}(t = 0)$ and $\theta\textsubscript{body}(t = 0)$. We then calculated the falling time for the rigid robot to rotate from an initial body orientation to the final upright orientation ($\theta\textsubscript{body} \approx 90^\circ$).

\subsection{Falling time}

\begin{figure}[h]
      \centering
      \includegraphics[scale=0.8]{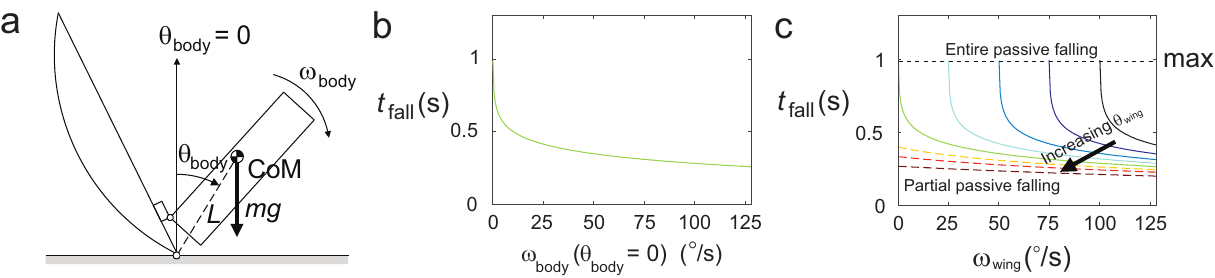}
      \caption{
      Planar dynamic model of the robot during the falling phase of winged self-righting. (a) Schematic of the robot as a rigid body falling under gravity about a fixed pivot on the ground. (b) Falling time, $t\textsubscript{fall}$, as a function of body pitch angular velocity when the body was vertical, $\omega\textsubscript{body}(\theta\textsubscript{body} = 0)$, predicted from the model for the case of entire passive falling. (c) $t\textsubscript{fall}$ as a function of $\omega\textsubscript{wing}$ for a wide range of $\theta\textsubscript{wing}$. Solid: entire passive falling. Dashed: partial passive falling. Arrow indicates direction of increasing $\omega\textsubscript{wing}$.
      }
      \label{dynamics}
\end{figure}

The behavior of the robot during the falling phase could be divided into two cases. The first case (entire passive falling) occurred when wing opening magnitude was small enough (small $\theta\textsubscript{wing}$) that wing opening had already stopped when the robot body reached a vertical orientation ($\theta\textsubscript{body}$ = 0). In this case, the entire falling phase was passive falling under gravity. Thus, we calculated falling time using the initial condition $\theta\textsubscript{body}$($t$ = 0) = 0. The second case (partial passive falling) occurred when wing opening was large enough (large $\theta\textsubscript{wing}$) such that, even after the robot body reached a vertical orientation ($\theta\textsubscript{body}$ = 0), the wings continued to open during an initial stage of the falling phase. In this case, passive falling only occurred during part of the falling phase. Thus, we calculated falling time using initial condition $\theta\textsubscript{body}$($t$ = 0) $>$ 0. For the entire passive falling case, we found from the model that $t\textsubscript{fall}$ decreased monotonically with the angular velocity of the body when the body was vertical, $\omega\textsubscript{body}(\theta\textsubscript{body} = 0)$ (Figure~\ref{dynamics}b). For the partial passive falling case, we found that $t\textsubscript{fall}$ not only decreased monotonically with $\omega\textsubscript{body}(t = 0)$, but also had a lower upper bound at $\omega\textsubscript{body}(t = 0) = 0$ that decreased monotonically with $\theta\textsubscript{body} (t = 0)$.

Our model predictions of $t\textsubscript{fall}$ vs. $\omega\textsubscript{body}$ provided us with a means to explain the observed dependence of falling time on $\theta\textsubscript{wing}$ and $\omega\textsubscript{wing}$ (Figure~\ref{performance}d, Figure~\ref{performanceweight}d). This is because, for any given $\theta\textsubscript{wing}$, we observed that increasing $\omega\textsubscript{wing}$ resulted in an approximately proportional increase of $\omega\textsubscript{body}$ at the end of wing opening (or beginning of passive falling). This meant that the observed $t\textsubscript{fall}$ vs. $\omega\textsubscript{wing}$ reflected how $t\textsubscript{fall}$ depended on $\omega\textsubscript{body}$ ($t$ = 0) for any given $\theta\textsubscript{wing}$ (apart from scaling the $\omega\textsubscript{wing}$-axis). As $\theta\textsubscript{wing}$ was varied, it was only possible at one particular $\theta\textsubscript{wing}$ for wing opening to stop exactly when the body was vertical. For this $\theta\textsubscript{wing}$ (Figure~\ref{dynamics}c, green curve), $t\textsubscript{fall}$ vs. $\omega\textsubscript{wing}$ had the same shape as $t\textsubscript{fall}$ vs. $\omega\textsubscript{body}(\theta\textsubscript{body} = 0)$ (apart from scaling the $\omega\textsubscript{wing}$-axis). As $\theta\textsubscript{wing}$ was reduced, the wings had stopped before the body reached the vertical orientation. The falling phase was still entirely passive, but by the time the body reached the vertical orientation, $\omega\textsubscript{body}$ must have reduced by a constant. In this case, $t\textsubscript{fall}$ vs. $\omega\textsubscript{wing}$ had the same shape as $t\textsubscript{fall}$ vs. $\omega\textsubscript{body}(\theta\textsubscript{body} = 0)$ but shifted to the right (Figure~\ref{dynamics}c, cyan and blue curves). As $\theta\textsubscript{wing}$ was increased, the robot started passive falling at $\theta\textsubscript{body} > 0$ and the falling phase was only partially passive. In this case, $t\textsubscript{fall}$ vs. $\omega\textsubscript{wing}$ had the same shape as $t\textsubscript{fall}$ vs. $\omega\textsubscript{body}$ as predicted for the partial passive falling case (Figure~\ref{dynamics}c, orange and red curves). Overall, our model predictions of $t\textsubscript{fall}$ vs. $\omega\textsubscript{wing}$ and $\theta\textsubscript{wing}$  (Figure~\ref{dynamics}b) qualitative matched experimental observations (Figure~\ref{performance}d, Figure~\ref{performanceweight}d), with an accurate upper bound of 1 second.

Using the inertial properties and kinematic data, we also estimated that the robot needed mechanical power of a range of 0.03 to 0.22 W when wing pitch angular velocity varied from 25 to 125 $^\circ$/s.

These results provided insights into how the dynamics of winged self-righting depended on wing opening speed and magnitude. When wing opening magnitude was given, the faster wings opened and pushed against the ground, the more kinetic energy it could inject into the body, resulting in faster falling. The smaller the wing opening, the earlier the robot stopped pushing against the ground. Thus, the kinetic energy gained during wing opening must be used to further raise the center of mass to the highest position before passive falling started. As wing opening magnitude increased beyond that required to raise the center of mass to the highest position, the extra more kinetic energy then accelerated the falling process.

Our simple dynamic model also provided insights into the dynamics of the discoid cockroach's winged self-righting. In animal experiments~\cite{li2015fast}, we observed that the discoid cockroach's wing opening stopped well before its body reached the highest center of mass position. Therefore, its falling phase was entirely passive. Using our model with physical parameters from the animal~\cite{li2015fast}, we found that it took the discoid cockroach significantly less time to fall to the ground than predicted from the model, assuming that passive falling began with no initial kinetic energy. This means that the animal could gain more than enough kinetic energy to not only overcome potential energy barriers for self-righting, but also speed up the falling process. Thus, winged self-righting of the discoid cockroach is highly dynamic. Further, the result that deficiencies associated with small wing opening magnitudes could be compensated by faster wing opening may provide insights into the trade-offs of prioritizing muscle contraction magnitude vs. velocity, both of which strongly affect muscle force production~\cite{Dickinson2000animals}.

\section{SYMMETRIC RIGHTING GEOMETRIC MODELING}

\subsection{Geometric model of the wings}

\begin{figure}[b]
      \centering
      \includegraphics[scale=0.81]{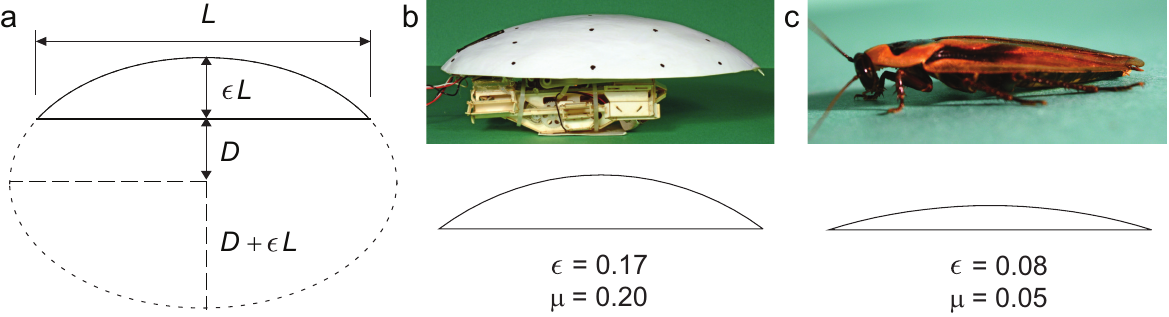}
      \caption{
      Geometric model of the robot self-righting using symmetric wing opening. (a) Parameterization of the truncated ellipse using normalized wing height $\epsilon$, defined as the ratio of wing height to wing length $L$ for a given truncation distance $D$. (b) Side view of the robot with $\epsilon$ = 0.17 and $\mu$ = 0.20 (for total mass $m$ = 100 g). (c) Side view of the discoid cockroach with $\epsilon$ = 0.08 and $\mu$ = 0.05.
      }
      \label{geometrymodel}
\end{figure}

To understand the energetic requirements for self-righting using symmetric wing opening and to determine whether the robot could self-right quasi-statically or must right dynamically, we utilized a generic planar self-righting analysis framework~\cite{Kessens2013Framework} and its associated self-rightability metric~\cite{Kessens2014metric}. We approximated the body shape in the sagittal plane as a rectangle and the wing shape in the sagittal plane as a rigid truncated ellipse (Figure~\ref{geometrymodel}a). The truncated ellipse was a reasonable approximation because wing rolling motion only slightly changed the wing shape in the sagittal plane except when $\theta\textsubscript{wing} > 60 ^\circ$, at which time the ground contact point was near the front edge of the wings and depended little on wing shape.

To study how wing shape affects self-righting, we fixed the wing length ($L$ = 18 cm) and the truncation distance ($D$ = 13 cm) while allowing the wing height $\epsilon L$ to vary, where $\epsilon$ is wing height normalized to wing length (Figure~\ref{geometrymodel}a). A small $\epsilon$ yielded thinner wings, while a larger $\epsilon$ yielded taller wings. For example, the robot has relatively taller wings of $\epsilon$ = 0.17 (Figure~\ref{geometrymodel}b) than the discoid cockroach's wings of $\epsilon$ = 0.08 (Figure~\ref{geometrymodel}c). To study how the mass distribution between the body and the wings affected self-righting, we also varied relative wing mass $\mu = m\textsubscript{wings}/(m\textsubscript{body} + m\textsubscript{wings})$, the ratio of wing mass to total mass. The robot had a relative wing mass of $\mu$ = 0.20 for a total mass of $m$ = 100 g (or 0.24 for $m$ = 83 g), and the discoid cockroach had a relative wing mass of $\mu$ = 0.05.

\subsection{Quasi-static righting analysis}

\begin{figure}[thpb]
      \centering
      \includegraphics[scale=0.8]{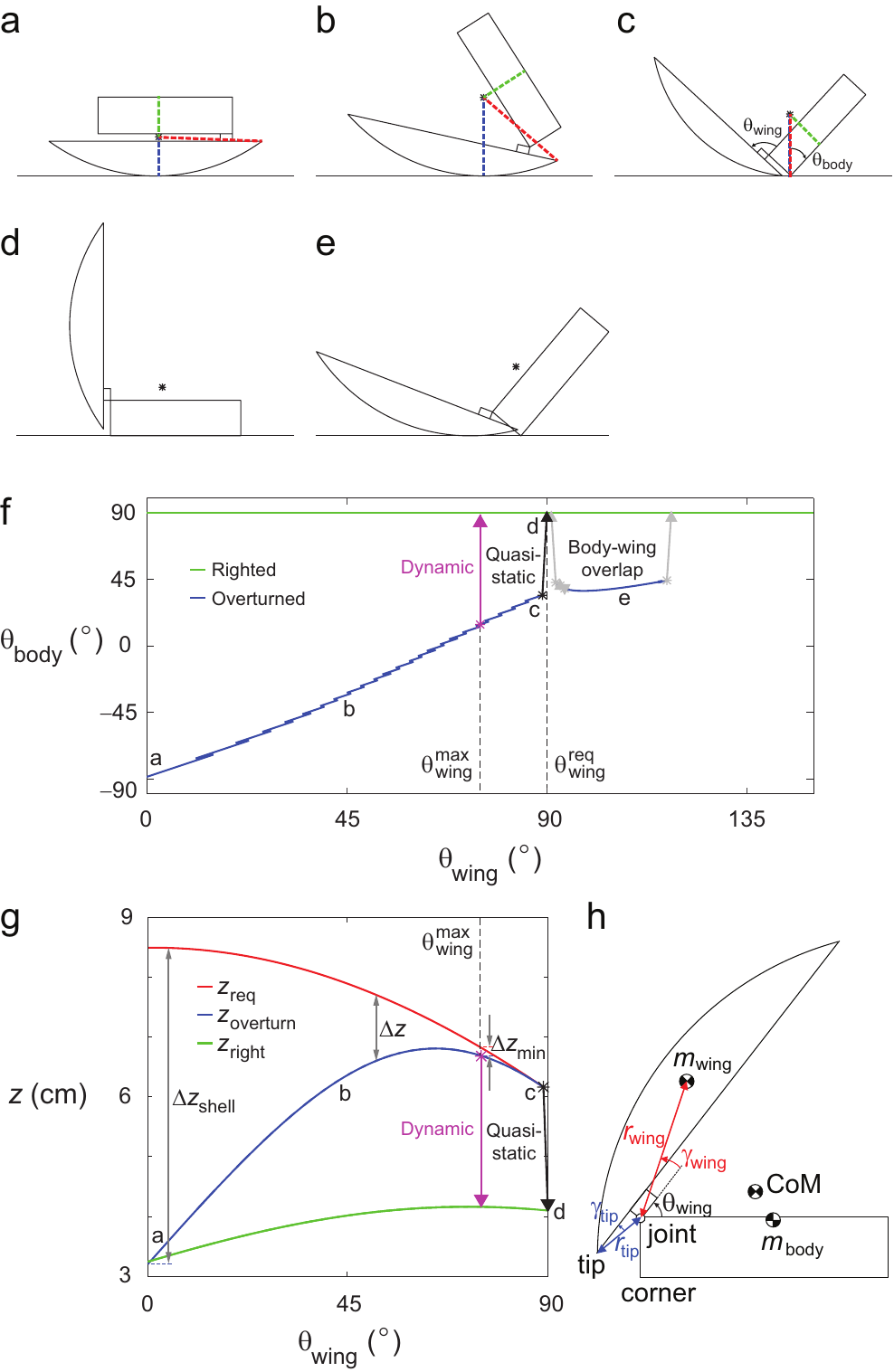}
      \caption{
      Robot quasi-static stability analysis. (a-e) Side views of the robot geometric model at representative stable body orientations as the wings open. (f) Robot body orientation $\theta\textsubscript{body}$ as a function of wing pitch angle $\theta\textsubscript{wing}$ during quasi-static wing opening. This is also known as the stable state map of the robot. (g) Robot center of mass height $z$ as a function of wing pitch angle $\theta\textsubscript{wing}$ calculated from the geometric model. (h) Definitions of geometric parameters used to derive analytical solutions of the red and green curves in (g). In (f) and (g), blue and green curves correspond to stable overturned body orientations and the stable righted body orientations, respectively. Black arrow shows a self-righting transition that is achievable with quasi-static wing opening. Magenta arrow shows a transition using dynamic wing opening where the potential energy barrier is lowest for $\theta\textsubscript{wing} \leq$ 75$^\circ$, the maximal wing pitch angle achievable by the physical robot. In (g), red curve represents the minimum center of mass height that would need to be achieved to self-right. The difference between the blue and red curves, $\Delta z$, is the potential energy barrier height that the robot must overcome. $\Delta z$ is highest for a static shell ($\theta\SPSB{max}{wing}$ = 0), and is lowest when wing opening was maximal ($\theta\SPSB{max}{wing}$ = 75$^\circ$ for the physical robot). Blue, red, and green dashed lines in (a)-(c) correspond with the blue, red, green curves in (g) at the corresponding instances.
      }
      \label{framework}
\end{figure}

The geometric self-righting framework~\cite{Kessens2013Framework} allowed us to understand how the body rotated as the wings opened quasi-statically by analyzing how the center of mass shifted relative to ground contact (Figure~\ref{framework}). The robot was initially placed upside down on level ground ($\theta\textsubscript{body}$ = $-$90$^\circ$), with the wings folded against the body ($\theta\textsubscript{wing}$ = 0) (Figure~\ref{framework}a). As the wings opened (increasing $\theta\textsubscript{wing}$), the center of mass position shifted towards the front end of the wings (Figure~\ref{framework}b). Because for quasi-static self-righting, the center of mass must always be directly above the single ground contact point of the wings that locally minimizes the robot's potential energy, the robot then rotated clockwise ($\theta\textsubscript{body}$ increased) (Figure~\ref{framework}f, blue curve). When the center of mass fore-aft position shifted beyond the front edge of the wings (Figure~\ref{framework}c), the robot then fell to the ground (Figure~\ref{framework}f, black arrow) to reach the final upright orientation ($\theta\textsubscript{body}$ = 90$^\circ$) (Figure~\ref{framework}d). This static stability analysis showed that, as long as the wings opened by more than a threshold $\theta\SPSB{req}{wing}$, the robot was guaranteed to self-right quasi-statically. If the wings opened quasi-statically by less than $\theta\SPSB{req}{wing}$, the robot would not successfully right but stop in a stable body orientation of $\theta\textsubscript{body} <$ 90$^\circ$, in which case a dynamic maneuver using kinetic energy would be required.

Incidentally, the model of the physical robot showed an additional region of stable states where the wings overlapped with the body (Figure~\ref{framework}e). From states located in this region, the robot could self-right either by opening its wings further or by closing its wings. However, due to kinematic constraints of the joint and potential wing-body interference, the physical robot would never be capable of achieving these states.

In addition, the framework~\cite{Kessens2013Framework} also allowed us to elucidate the kinetic energy requirements to induce self-righting as the wings open (Figure~\ref{framework}g), by analyzing how the potential energy barrier height during self-righting depended on wing opening magnitude. We calculated the center of mass height for the robot ($\epsilon$ = 0.17, $\mu$ = 0.2) as a function of wing pitch angle when the robot was in its stable overturned orientation (Figure~\ref{framework}g, blue curve) and when the robot was in its stable righted orientation (green curve), as well as the farthest distance from the center of mass to any hull point along its rolling path to the righted orientation (red curve).

While these curves were generated using a numerical simulation of the robot starting in an overturned orientation, we could also derive them analytically. The green ``righted" curve, $z\textsubscript{right}$, was simply the minimum distance from the robot's center of mass (CoM) to the bottom of the body, representing the height of the center of mass in the robot's righted orientation. This was described by:

\begin{equation}
z\textsubscript{right}= z\textsubscript{CoM} = \frac{m\textsubscript{body} z\textsubscript{body} + m\textsubscript{wing} [z\textsubscript{joint} + r\textsubscript{wing} \text{sin}(\theta\textsubscript{wing} + \gamma\textsubscript{wing})]}{m\textsubscript{body} + m\textsubscript{wing}}
\end{equation}

The red curve, $z\textsubscript{req}$, represented the maximum center of mass height as the robot rolled from the overturned orientation to the righted orientation. For the specific geometry of the robot in this paper, this was represented by the distance from the robot's center of mass (CoM) either to the front wing tip or to the front bottom corner of the body. This was described by:

\begin{equation}
    z\textsubscript{req} = \text{max}
    \begin{cases}
      (x\SPSB{2}{CoM} + z\SPSB{2}{CoM})^{\frac{1}{2}} \\
      [(x\textsubscript{CoM} - x\textsubscript{tip})^2 + (z\textsubscript{CoM} - z\textsubscript{tip})^2]^{\frac{1}{2}}
    \end{cases}
\end{equation}\
where
\begin{equation}
x\textsubscript{CoM} = \frac{m\textsubscript{body} x\textsubscript{body} + m\textsubscript{wing} r\textsubscript{wing} \text{cos}(\theta\textsubscript{wing} + \gamma\textsubscript{wing})}{m\textsubscript{body} + m\textsubscript{wing}}
\end{equation}
\begin{equation}
x\textsubscript{tip} = - r\textsubscript{tip} \text{cos}(\theta\textsubscript{wing} - \gamma\textsubscript{tip})
\end{equation}
\begin{equation}
z\textsubscript{tip} = z\textsubscript{joint} - r\textsubscript{tip} \text{sin}(\theta\textsubscript{wing} - \gamma\textsubscript{tip})
\end{equation}

The blue curve, $z\textsubscript{overturn}$, was more difficult to express analytically, because it represented the minimal distance from the center of mass to the portion of the convex hull in contact with the ground as the robot rotated from the overturned state and the righted state. The point on the convex hull at which this occurred changed for each wing pitch angle. Thus, we continued to rely on our simulation to generate the blue curve. Definitions of the geometric parameters used to derive these analytical solutions are shown in Figure~\ref{framework}h.

The difference between the blue and red curves, $\Delta z$, represented the potential energy barrier height that must be overcome in order to self-right, which depended on wing pitch angle. If the robot was able to open its wings by a large enough magnitude ($\theta\textsubscript{wing} \geq \theta\SPSB{req}{wing}$), then it could right quasi-statically (Figure~\ref{framework}g, black arrow) and did not require kinetic energy to overcome potential energy barriers (because $\Delta z\textsubscript{min}$ vanished). However, if the robot was not able to do so ($\theta\textsubscript{wing} < \theta\SPSB{req}{wing}$), it must self-right dynamically (Figure~\ref{framework}g, magenta arrow) by injecting sufficient kinetic energy to overcome the potential energy barrier $\Delta z$. The potential energy barrier was highest for $\theta\textsubscript{wing} = 0$, the case where the wings were reduced to a static shell~\cite{Li2015terradynamically}.

\subsection{Effect of wing shape \& mass on quasi-static self-rightability}

\begin{figure}[h]
      \centering
      \includegraphics[scale=1.23]{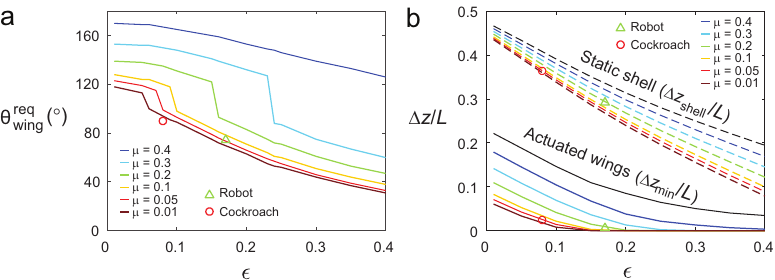}
      \caption{
      Dependence of kinematic and energetic requirements of winged self-righting on wing shape and relative wing mass. (a) Minimal wing pitch angle required, $\theta\SPSB{req}{wing}$, for quasi-static self-righting as a function of normalized wing height $\epsilon$ for a wide range of relative wing mass $\mu$. (b) Minimum potential energy barrier height $\Delta z$ (normalized by wing length $L$) that the robot must overcome to self-right from the stable overturned orientation with wings closed as a function of $\epsilon$ for a wide range of $\mu$. Solid: if the robot can open wings to $\theta\SPSB{max}{wing}$ = 75$^\circ$. Dashed: if the wings cannot be opened, i.e., if they are a static shell. These two cases correspond with $\Delta z\textsubscript{min}$ and $\Delta z\textsubscript{shell}$ in Figure~\ref{framework}b, respectively. In (a) and (b), green triangle and red circle represent model predictions for the physical robot ($\epsilon$ = 0.17 and $\mu$ = 0.20 for total mass $m$ = 100 g) and the discoid cockroach ($\epsilon$ = 0.08 and $\mu$ = 0.05), respectively.
      }
      \label{geometryresults}
\end{figure}

To understand how wing shape and mass distribution affect the kinematic requirement for dynamic self-righting, we varied relative wing height and relative wing mass and studied how they affected the minimum wing pitch angle required for quasi-static self-righting (Figure~\ref{geometryresults}a). We found that, as long as the wings weighed less than the body ($\mu <$ 0.5), $\theta\SPSB{req}{wing}$ always decreased as the wings became taller (increasing $\epsilon$). For any given relative wing height (given $\epsilon$), $\theta\SPSB{req}{wing}$ was always larger when the wings were heavier (larger $\mu$). When the wings became as heavy as or heavier than the body ($\mu \geq$ 0.5), the robot was unable to right itself quasi-statically, because the lighter body lifted off the ground instead of the heavier wing. The sudden increase of $\theta\SPSB{req}{wing}$ when $\epsilon$ became small was due a topological change in the stable state space map. This occurred when the region containing states where the wings overlapped with the body (Figure~\ref{framework}e) merged with the region containing states supported solely by the wings (Figure~\ref{framework}d). However, the potential energy barrier that must be overcome in the region of the transition remains small, indicating that dynamic righting would be easier at these wing pitch angles.

Together, these observations demonstrated that a robot (or animal) with taller and lighter wings can more easily self-right quasi-statically, which may be necessary if it is incapable of opening wings quickly (for example, if the wing joints were highly geared). The result that $\theta\SPSB{req}{wing}$ decreased with $\mu$ was also consistent with the experimental observation that, at the slowest wing opening tested ($\omega\textsubscript{wing}$ = 25$^\circ$/s), the robot with relatively lighter wings was able to right at a smaller wing opening magnitude ($\theta\textsubscript{wing}$ = 55$^\circ$ for $\mu$ = 0.20) than that with relatively heavier wings ($\theta\textsubscript{wing}$ = 58$^\circ$ for $\mu$ = 0.24).

These results allowed us to determine whether our robot was able to self-right quasi-statically or must use kinetic energy to right dynamically. For the robot used in experiments ($\epsilon$ = 0.17, $\mu$ = 0.20 for $m$ = 100 g), the model predicted that it must open its wings to $\theta\SPSB{req}{wing} \approx 90^\circ$, beyond the maximal wing pitch angle $\theta\SPSB{max}{wing}$ = 75$^\circ$ that the robot is capable of (Figure~\ref{geometryresults}a, green triangle). This means that the robot could not self-right quasi-statically and must use kinetic energy to right dynamically. We verified that the slight deformation of wing shape due to the roll motion of the wings at large $\theta\textsubscript{wing}$ did not change this prediction.

Similarly, we could also determine whether the discoid cockroach could quasi-statically self-right. The discoid cockroach has thin wings ($\epsilon$ = 0.08) and small relative wing mass ($\mu$ = 0.05), which results in $\theta\SPSB{req}{wing} \approx$ 100$^\circ$. By contrast, our animal observations showed that the discoid cockroach could only open its wings by a maximum of $\theta\SPSB{max}{wing}\approx$ 90$^\circ$ (Figure~\ref{geometryresults}a, red circle) during self-righting~\cite{li2015fast}. This suggested that the animal uses kinetic energy to self-right dynamically.

\subsection{Effect of wing shape \& mass on potential energy barrier}

To understand how wing shape and mass distribution affect the energetic requirement for dynamic self-righting, we determined the minimal potential energy barrier height, $\Delta z\textsubscript{min}$, that must be overcome by using kinetic energy for the wide range of relative wing height and relative wing mass tested (Figure~\ref{geometryresults}b, solid curves). We constrained wing pitch angle to be 0 $\leq \theta\textsubscript{wing} \leq$ 75$^\circ$, as the robot could only open its wings to $\theta\SPSB{max}{wing}$ = 75$^\circ$, and the discoid cockroach also rarely opened its wings beyond this range~\cite{li2015fast}. We found that, for any given $\mu$, $\Delta z\textsubscript{min}$ decreased monotonically with $\epsilon$. This was because for the range of $\epsilon$ tested, taller wings were closer to semi-spherical, for which the potential energy barrier height diminished. In addition, $\Delta z\textsubscript{min}$ was always higher for larger $\mu$ for any given $\epsilon$. This was because as the wings became heavier relative to the body, center of mass height became lower when the robot was upside down but did not significantly change when the robot was near vertical, thus increasing the potential energy barrier height. Therefore, tall, lightweight wings are energetically advantageous for winged dynamic self-righting.

To study how much energy winged self-righting helped the robot and animal save compared to if they do not use wings (such as turtles~\cite{Domokos2008Geometry}), we further calculated the potential energy barrier height, $\Delta z\textsubscript{shell}$, with a static shell ($\theta\SPSB{max}{wing}$ = 0) (Figure~\ref{geometryresults}b, dashed curves). We found that $\Delta z\textsubscript{shell}$ also decreased monotonically with $\epsilon$ and was higher for larger $\mu$ for a static shell, consistent with the case of using wings. However, $\Delta z\textsubscript{shell}$ was much higher for a static shell than for using wings to self-right: for example, by a factor of 2 for $\mu$ = 0.5 and an order of magnitude for $\mu$ = 0.05. This demonstrated that, while both the discoid cockroach and our winged robot are incapable of self-righting quasi-statically using wings, their ability to alter overall shape by opening wings allows them to use significantly less kinetic energy to dynamically self-right. Further, the mechanical energy needed to open the wings quasi-statically equals the work needed to raise the center of mass to its maximal height during quasi-static wing opening (Figure~\ref{geometryresults}g, peak of the blue curve). Therefore, winged dynamic self-righting is more economical than passive self-righting using rigid shells (e.g., turtles with highly-domed shells~\cite{Domokos2008Geometry}).

It is worth noting that, while taller wings may be energetically advantageous for winged dynamic self-righting, other functions such as protection and obstacle traversal~\cite{Li2015terradynamically} also play an important role in the evolution of the wing shape. For robots to begin to achieve animal-like multi-functionality~\cite{Low2015Perspectives}, we must also consider such design trade-offs.

\section{ASYMMETRIC RIGHTING EXPERIMENTS}

\subsection{Biological hypothesis}

\begin{figure}[thpb]
      \centering
      \includegraphics[scale=0.86]{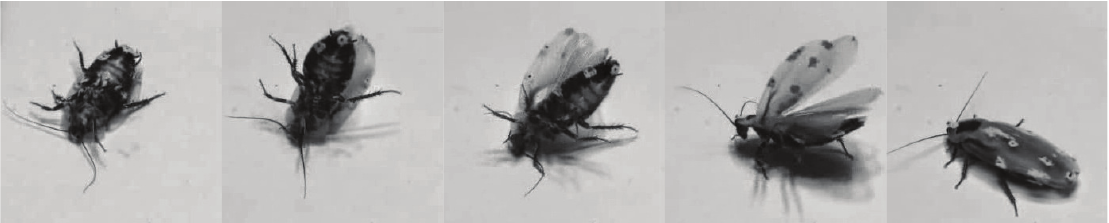}
      \caption{
      Self-righting of the discoid cockroach using asymmetric wing opening and body rotation out of the sagittal plane.
      }
      \label{hypothesis}
\end{figure}

In our animal experiments~\cite{li2015fast}, we observed that, when the discoid cockroach used wings to self-right, it often took the animal multiple maneuvers to eventually right itself. In addition, while the animal is capable of opening its wings to achieve wing pitch angles as large as $\approx$ 90$^\circ$, it most often opened them by much smaller magnitudes even during successful self-righting maneuvers. Further, the animal rarely opened the two wings symmetrically and righted by pitching over its head in the sagittal plane; instead, it usually opened the two wings asymmetrically and righted using body rotations out of the sagittal plane (Figure~\ref{hypothesis}). Thus, we hypothesized that it may be advantageous to use asymmetric wing opening to self-right.

Our robot provided us with a platform to test this hypothesis thanks to its ability to open the two wings independently. We systematically tested the robot (total mass $m$ = 100 g) by opening the left and right wings by to different wing pitch angles, $\theta_L$ and $\theta_R$, at different speeds, $\omega_L$ and $\omega_R$. Opening of the two wings always started and stopped in synchrony with no phase lag, i.e., $\theta_L$/$\omega_L$ = $\theta_R$/$\omega_R$, with the wing that opened by the a larger magnitude always opening at 125$^\circ$/s. We performed 10 trials for each combination of $\theta_L$ and $\theta_R$ because we observed larger trial-to-trial variations for asymmetric wing opening than for symmetric wing opening.

\subsection{Righting outcomes \& probability}

\begin{figure}[b]
      \centering
      \begin{tabular}
        {@{}ccc@{}}
            \includegraphics[scale=0.88]{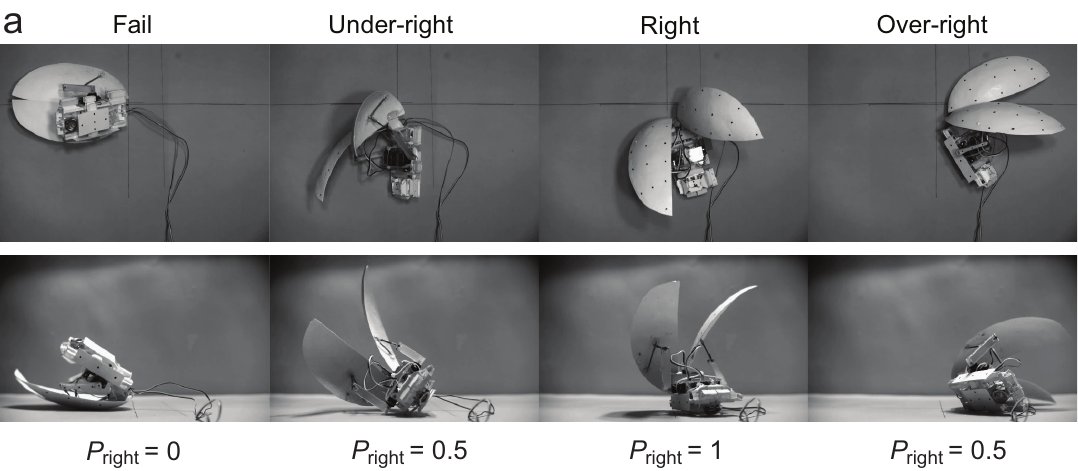} \\
            \includegraphics[scale=0.57]{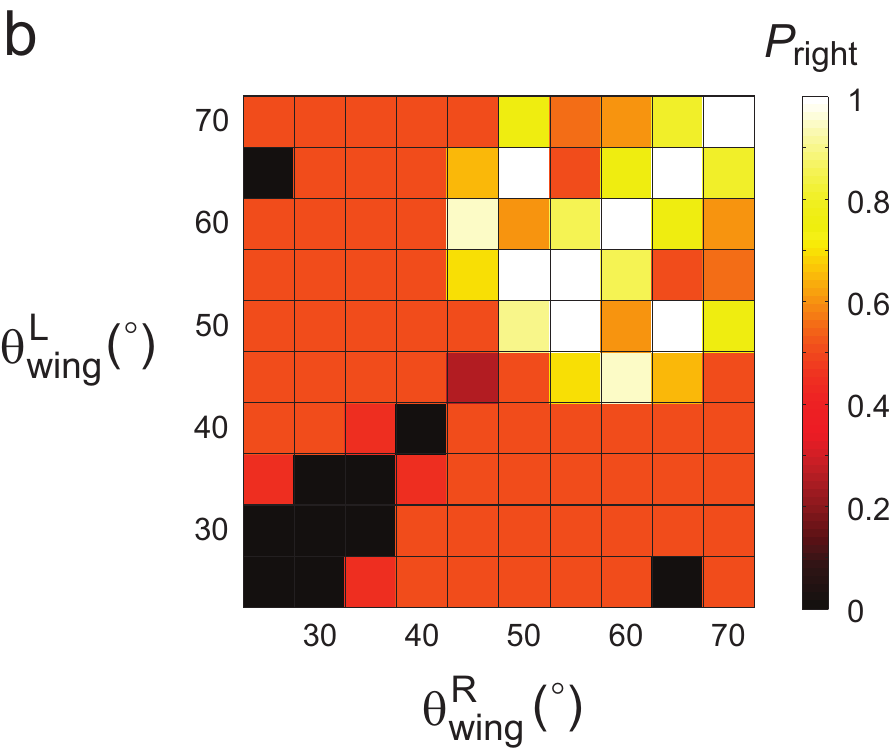} \\
        %\multicolumn{2}{c}{\includegraphics[scale=0.55]{./Figures/Fig10.2.eps}}
      \end{tabular}
      \caption{
      Self-righting of the robot using asymmetric wing opening. (a) Top and side views of the robot's final body orientation for the four outcomes observed: failure, under-righting, successful righting, and over-righting. (b) Average righting probability, $P\textsubscript{right}$, as a function of left and right wing opening magnitudes, $\theta_L$ and $\theta_R$.
      }
      \label{asymmetric}
\end{figure}

We observed that, with asymmetric wing opening, the robot body rotation was not purely pitching within the sagittal plane, but also had a rolling component. This led to more diverse righting outcomes than using symmetric wing opening (Figure~\ref{asymmetric}a). When $\theta_L$ and $\theta_R$ were both small and not too different, the robot sometimes failed to right, settling into a stable orientation with the body slightly pitched up similar to failure for symmetric wing opening, except that the body had also rolled slightly. As $\theta_L$ and $\theta_R$ increased, the robot body was able to move out of this stable orientation. As the body fell to the ground, due to the asymmetric body rotation, it usually first contacted the ground not on its ventral surface as in the case of symmetric wing opening, but instead on the edge of its side. Depending on how much kinetic energy the robot had during the body-ground collision, the body may or may not be able to continue to roll onto its ventral surface to reach an upright orientation, resulting in either under-righting or successful righting. Occasionally, the body still had substantial kinetic energy left after it rolled to the upright orientation, and continued to roll and eventually fell onto its other side, resulting in over-righting. In animal experiments~\cite{li2015fast}, we also observed under-righting and over-righting when the discoid cockroach used wings to self-right. However, the animal always achieved the upright orientation afterwards using its legs.

To examine whether asymmetric wing opening enhanced the probability of the robot self-righting, we defined failure, under-righting, successful righting, and over-righting to have righting probability of $P\textsubscript{right}$ = 0, 0.5, 1, and 0.5, respectively, and calculated average righting probability for each combination of $\theta_L$ and $\theta_R$ tested. We found that for large $\theta_L$ and $\theta_R$ ($>$ 50$^\circ$), $P\textsubscript{right}$ was usually high ($>$ 80\%) and insensitive to wing opening asymmetry (Figure~\ref{asymmetric}b, top right quarter). Surprisingly, for small $\theta_L$ and $\theta_R$ ($\leq$ 45$^\circ$), asymmetric wing opening consistently resulted in higher $P\textsubscript{right}$ ($\approx$ 50\%) than symmetric wing opening did ($\approx$ 0) (Figure~\ref{asymmetric}b, bottom left quarter). This suggests that, when wing opening magnitudes are limited (e.g., when an animal fatigues or when a robot has low power), asymmetric wing opening is more advantageous by increasing the chance of righting.

\subsection{Heading change}

\begin{figure}[b]
      \centering
      \begin{tabular}
        {@{}ccc@{}}
            \includegraphics[scale=0.77]{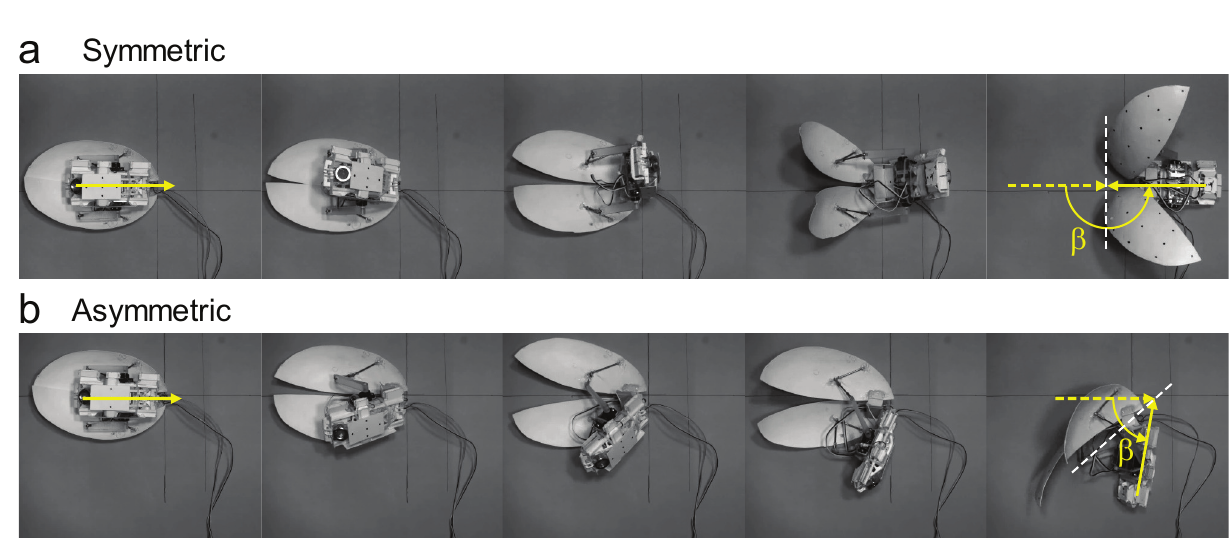} \\
            \includegraphics[scale=0.57]{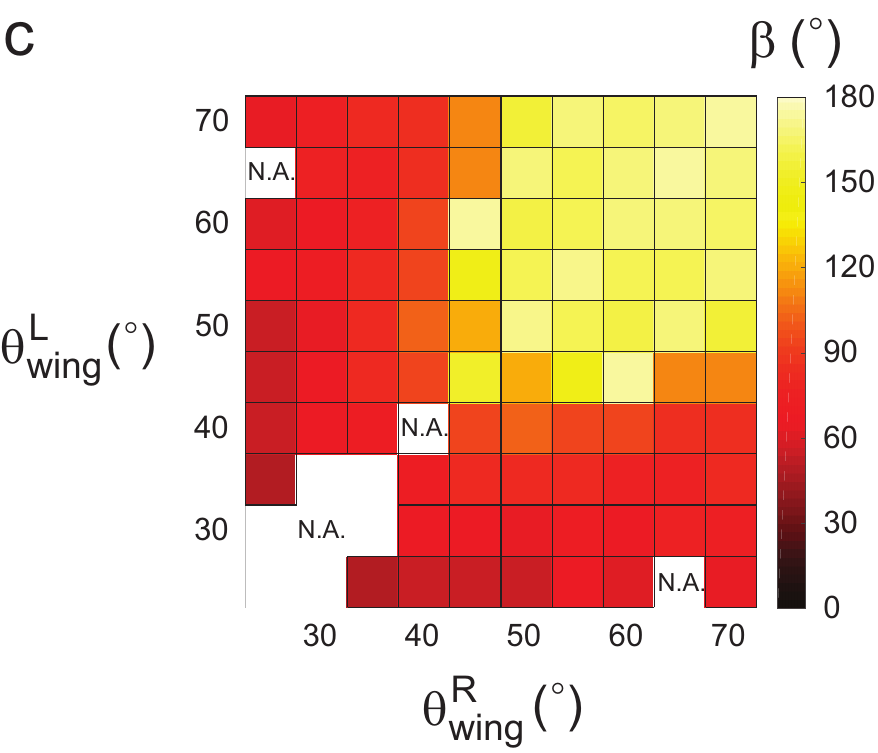} \\
      \end{tabular}
      \caption{
      Change of robot heading from before to after the self-righting maneuver. (a) Self-righting using symmetric wing opening always resulted in change of heading of $\beta$ = 180$^\circ$. (b) Change of heading was reduced, $\beta <$ 180$^\circ$, as wing opening became asymmetric. In (a) and (b), yellow arrows represent robot heading, and white dashed line represents approximate body rotation axis within the ground plane. (c) Average change of heading, $\beta$, as a function of left and right wing opening magnitudes, $\theta_L$ and $\theta_R$.
      }
      \label{orientation}
\end{figure}

Because the robot body rotation was out of the sagittal plane and had a rolling component during asymmetric righting, the highest position that its center of mass reached (and thus the potential energy the robot overcame) must be lower than that during symmetric wing opening. This suggests that the degree of asymmetric body rotation out of the sagittal plane could provide a means to infer the potential energy barrier for self-righting using asymmetric wing opening. Therefore, we measured the change in the heading of the robot, $\beta$, from the initial upside-down body orientation to the instant when the body fore-aft axis first became parallel to the ground after the robot fell to the ground (after which the axis of rotation would change). We observed that when wing opening was symmetric, the robot heading always changed by $\beta \approx$ 180$^\circ$ because the body always rotated about the body pitch axis within the sagittal plane (Figure~\ref{orientation}a). For asymmetric wing opening, $\beta$ was always smaller than 180$^\circ$ due to body rotation out of the sagittal plane (Figure~\ref{orientation}b). Changes in heading were usually greater than around 150$^\circ$ for large asymmetric wing opening ($\theta_L$ and $\theta_R \geq$ 50$^\circ$), and were often smaller than 90$^\circ$ for small asymmetric wing opening ($\theta_L$ and $\theta_R \leq$ 45$^\circ$) (Figure~\ref{orientation}c).

\section{ASYMMETRIC RIGHTING MODELING}

\subsection{Simple geometric model}

\begin{figure}[thpb]
      \centering
      \includegraphics[scale=0.55]{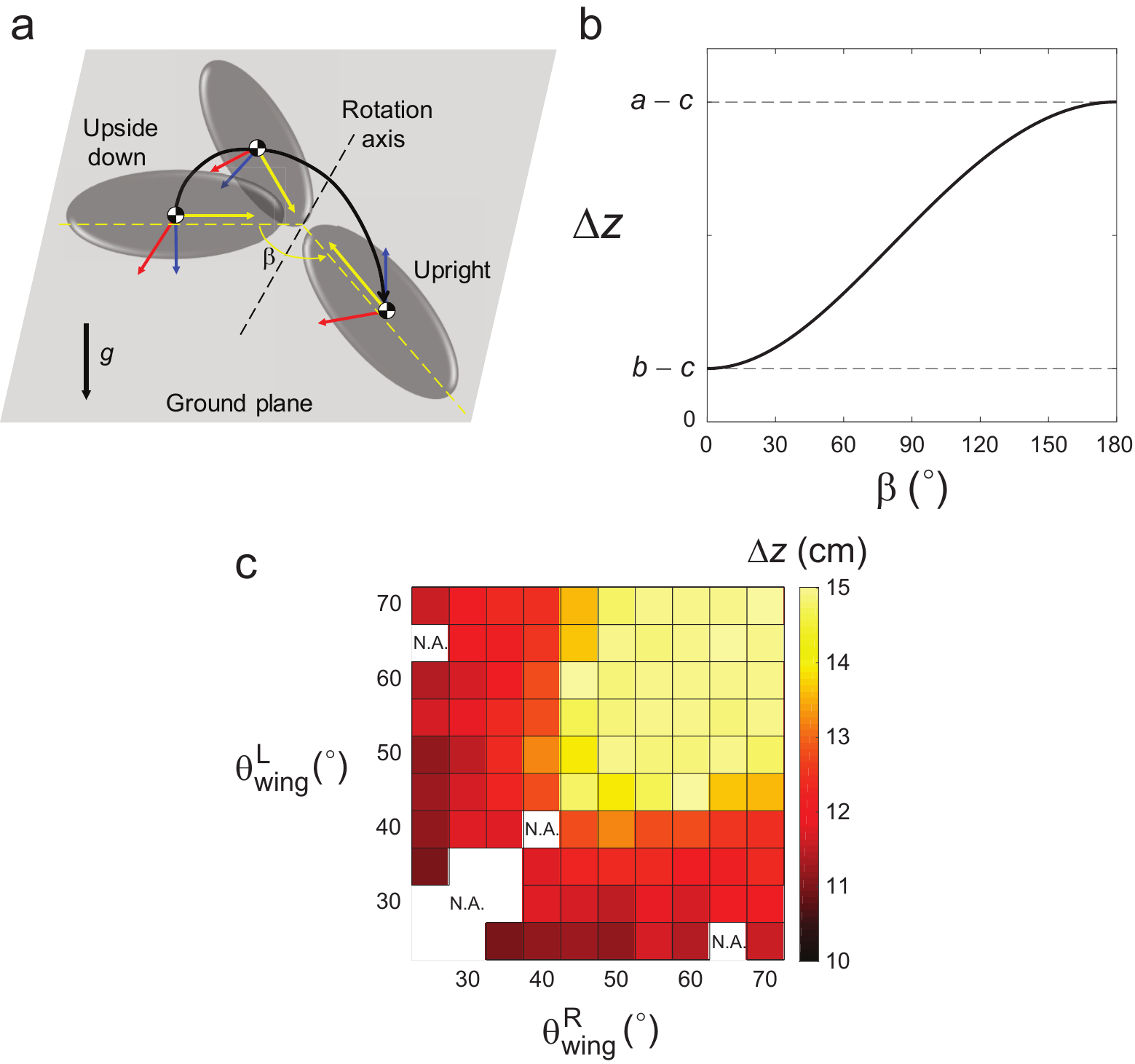}
      \caption{
      Simple geometric model of self-righting using asymmetric wing opening to estimate potential energy barrier height. (a) Schematic of an ellipsoidal body rotating about a fixed axis within the ground plane, resulting in change of heading by $\beta$. Yellow, red, and blue arrows represent the fore-aft, lateral, and dorso-ventral directions. (b) Potential energy barrier height, $\Delta z$, as a function of $\beta$ predicted from the model. $a$, $b$, and $c$ are the length of the semi axes of the ellipsoid in the fore-aft, lateral, and dorso-ventral directions, respectively. (c) $\Delta z$ as a function of left and right wing opening magnitudes, $\theta_L$ and $\theta_R$, calculated from the model using the average $\beta$ measured from asymmetric righting experiments.
      }
      \label{barrier}
\end{figure}

To estimate the potential energy barrier that the robot overcame during self-righting using asymmetric wing opening, we developed a simple geometric model of the robot. Although the robot's overall shape changed dynamically as the wings opened, examination of high speed videos showed that its body rotation during the self-righting process (Figure~\ref{orientation}a, b) could be approximated as rotation about a fixed axis within the ground plane to the first order. This axis was the pitch axis for symmetric wing opening (Figure~\ref{orientation}a), and was an axis between the initial body pitch and initial body roll axes for asymmetric wing opening (Figure~\ref{orientation}b). In addition, because the robot's rotation axis was always on the surface of the wings during the rising phase, the potential barrier that it overcame (which was proportional to the increase of center of mass height by the end of the rising phase) was primarily determined by the location of the rotational axis on the ellipsoidal wing. Therefore, we could simply treat the robot as an ellipsoid rotating about a fixed axis within the ground plane (Figure~\ref{barrier}a), with semi-axes of $a$ = 18 cm along the body fore-aft direction, $b$ = 13 cm along the body lateral direction, $c$ = 3 cm along the body dorso-ventral direction, and uniform mass distribution (i.e., center of mass was at geometric center).

\subsection{Potential energy barrier height}

Using this simple geometric model, we varied the angle between the rotation axis and the initial fore-aft axis of the body ($\beta$/2), and numerically calculated the potential energy barrier height (the increase of center of mass height during the rising phase), $\Delta z$, as a function of $\beta$, assuming that the lowest point of the body touched the ground. We found that $z\textsubscript{max}$ increased monotonically as the change of heading $\beta$ increased from 0 to 180$^\circ$ (Figure~\ref{barrier}b). This is intuitive because an elongate body ($a > b$) overcomes the highest potential energy barrier height $\Delta z = a - c$ when it pitches ($\beta$ = 180$^\circ$), overcomes the lowest potential energy barrier height $\Delta z = b - c$ when it rolls ($\beta$ = 0), and overcomes an intermediate potential energy barrier when it rotates about an axis between the initial pitch and roll axes.

Using the model-predicted dependence of center of mass height on change of heading and the average change of heading measured from the asymmetric righting experiments, we then estimated the center of mass height $z\textsubscript{max}$ for all the $\theta_L$ and $\theta_R$ tested (Figure~\ref{barrier}c). We found that $\Delta z$ decreased with the lower of the two the wing opening magnitudes and was lowest when the robot righted using asymmetric, small wing opening (e.g., $\theta_L$ = 25$^\circ$ and $\theta_R$ = 35 - 50$^\circ$). This suggests that, besides increasing righting probability, asymmetric wing opening also saves energy by reducing the potential energy barrier height when wing opening magnitudes are limited.

Our asymmetric righting experiments and modeling together suggest that the observed frequent occurrences of asymmetric wing opening~\cite{li2015fast} may be an adaptation of the discoid cockroach to right more quickly and economically. Unlike the robot whose wing opening could be precisely controlled to guarantee successful righting using the principles from our experiments, the discoid cockroach's wing opening can vary greatly from attempt to attempt. In the case an animal performs multiple consecutive unsuccessful maneuvers, its ability to open wings to large magnitudes and rapidly diminishes. Asymmetric wing opening can then greatly help the animal by not only increasing its probability of successful righting and reducing righting time, but also by saving the energy needed to right as a result of a lower potential energy barrier to be overcome and fewer unsuccessful attempts.

\section{CONCLUSIONS \& FUTURE WORK}

In this study, we began to understand the principles of dynamic terrestrial self-righting using wings through systematic experimentation and modeling of a novel winged robot inspired from the discoid cockroach. Analogous to exaptations~\cite{Jay1982exaptation} or co-opting of structures common in biological organisms, our novel winged self-righting mechanism provided rapidly-running small robots with a means to use existing body structures in novel ways to serve new locomotor functions~\cite{Mintchev2016adaptive}. Our symmetric righting experiments and modeling showed that, by opening its wings to large magnitudes and rapidly pushing against the ground, the robot can dynamically self-right using kinetic energy to overcome potential energy barriers. In addition, our model also suggested that the discoid cockroach's winged self-righting is a dynamic maneuver. Furthermore, our analysis of energetic requirement of self-righting showed that winged dynamic self-righting is more economical than passive righting using static shells. Finally, our asymmetric righting experiments and modeling showed that asymmetric wing opening is useful when the ability to open wings diminishes because it increases righting probability and reduces the potential energy barrier.

Inspired by our successful design and discovery of the principles of winged dynamic self-righting, we will continue to develop new robot prototypes that will have integrated ability to both traverse obstacles~\cite{Li2015terradynamically} and self-right if flipped over. In addition, we will continue to systematically test robots as physical models~\cite{Aguilar2016review} to further elucidate the mechanisms of dynamic self-righting, such as whether body vibrations induced by leg flailing help animals access lower energy barrier locomotor pathways~\cite{Li2015terradynamically}, how animals use legs to recover from under- and over-righting~\cite{li2015fast}, and how terrain topology~\cite{peng2015motion,Sasaki2016Reciprocity} and mechanics~\cite{Li2015terradynamically,Sasaki2016Reciprocity} affect self-righting. Further, measurements of ground reaction forces and multi-body dynamics simulations~\cite{li2013terradynamics} can provide more insights into the dynamics of the rising phase of winged self-righting. Finally, similar to how cockroaches can use mechanosensing to detect its change in body orientation to elicit self-righting response~\cite{Walthall1981receptors}, we will add an inertial measurement unit to the robot to detect flipping-over and initiate self-righting, and develop sensory feedback control strategies based on modeling insights to dynamically adjust body pitching and rolling for improved righting performance.

Together, the experimental and modeling frameworks that we are establishing will open doors to a more principled understanding of dynamic self-righting of a variety of animals~\cite{Brackenbury1990novel,full1995maximum,Frantsevich2004righting,Young2006effects,Domokos2008Geometry,li2015fast} and future robots~\cite{Saranli2004model,Saranli2004multipoint,Johnson2013legged,Li2016cockroach}. We envision that our integrative approach using biology to provide inspirations and hypotheses, robotics as physical models for systematic experiments, and physics principles to guide robot design and control will accelerate the advent of robots that can perform multi-functional locomotion~\cite{Low2015Perspectives} in complex terrains.

\section*{Acknowledgements}

We thank Austin Young for assistance with robot construction, motor calibration, and preliminary experiments; Rundong Tian, David Strachan-Olson, Zach Hammond, Will Roderick, Mel Roderick for help with early robot prototypes; Kaushik Jayaram, Nate Hunt, Tom Libby, Andrew Pullin for discussions; Ratan Othayoth for help with photos; and three anonymous reviewers for helpful comments.

This work is supported by Johns Hopkins University Whiting School of Engineering start-up funds (C.L.); Burroughs Wellcome Fund Career Award at the Scientific Interface (C.L.); Miller Institute for Basic Research in Science, University of California, Berkeley via a Miller Research Fellowship (C.L.); and United States Army Research Laboratory under the Micro Autonomous Systems and Technology Collaborative Technology Alliance (C.C.K., R.J.F).

Author contributions: C.L. conceived and designed study, performed experiments, developed dynamic model and asymmetric righting geometric model, analyzed data, and wrote the paper; C.C.K. designed study, developed symmetric righting geometric model, analyzed data, and wrote the paper; R.S.F. provided feedback on robot design; R.J.F. oversaw study.

\bibliographystyle{IEEEtran}

\label{lastpage}

\end{document}